\documentstyle{article}

\newcommand{\beq}[1]{\begin{equation} \label{#1} }
\newcommand{\eeq}   {\end{equation}}
\newcommand{\av}[1]{\langle #1 \rangle}

\begin{document}

\title{New Possibilities for Investigation of TRI Violation with
the use of Aligned Nuclei}
\author{V.A.Atsarkin$^1$, A.L.Barabanov$^2$, A.G.Beda$^3$,
V.V.Novitsky$^{3,4}$}
\date{}
\maketitle
\vspace{-5mm}

{\small

1. Institute of Radio Engineering and Electronics RAS, Moscow,
Russia

2. Kurchatov Institute, Moscow 123182, Russia

3. SSC Institute for Theoretical and Experimental Physics,
Moscow117259, Russia.

4. Joint Institute for Nuclear Research, Dubna 141980, Russia

}
\bigskip

\begin{abstract}
The methods of investigation of Time Reversal Invariance (TRI)
violation using polarized neutron beam and polarized or aligned
nuclear target are briefly considered. The new method of dynamic
nuclear alignment (DNA) of quadrupolar nuclei is proposed. An
implementation of this method can significantly increase the
number of aligned nuclei accessible for TRI violation experiments
and for the other physical investigations.
\end{abstract}

\section{Introduction}

Polarized neutron beams are the excellent tool for investigation
of fundamental symmetries violation, namely, parity (P) violation
and time reversal invariance (TRI) violation. At the moment a
great body of data is obtained on P violation in the interaction
of polarized neutrons with unpolarized nuclei. On the other hand,
a number of possible tests of TRI violation in similar experiments
with polarized and aligned targets are discussed.

The method to search for P- and T-odd nuclear interaction by
investigation of three-fold correlation
$({\bf n}_s[{\bf n}_k\times{\bf n}_I])$ in transmission of
polarized neutrons through polarized target was proposed in
\cite{1,2}. Here ${\bf n}_s$, ${\bf n}_I$ and ${\bf n}_k$ are unit
vectors along neutron and nuclear polarization axes and neutron
momentum, respectively. The possibility to search for P-even T-odd
nuclear interaction by studying of five-fold correlation
$({\bf n}_s[{\bf n}_k\times{\bf n}_I])({\bf n}_k{\bf n}_I)$ was
first considered in \cite{3}-\cite{5}. This method needs an
aligned target (note, that here ${\bf n}_I$ is a unit vector along
an alignment axis).

A difficulty of nuclear alignment presents a major obstacle for
five-fold correlation experiment. Up to now the number of nuclei
which were aligned is very limited. We propose the new method of
dynamic nuclear alignment (DNA) which can be used to increase
significantly the number of nuclei accessible for appropriate
physical experiments.

\section{TRI tests}

Both three- and five-fold correlation experiments are unique null
tests of TRI. The usually used methods as comparison of the cross
sections of direct and inverse reactions or of the polarization
and analyzing power in scattering experiments (see, e.g.,
\cite{6,7}) are based on measurements of two values, which should
coincide if TRI holds. Clearly, a measurement of a single value,
which is nonzero only if TRI breaks, is much more reliable.
Three- and five-fold correlations arise in the forward scattering
amplitude $f(0)$. Thus, they appear in the total cross section,
which is linear on $f(0)$ as a result of the optical theorem
\beq{1}
\sigma_{tot}=\frac{4\pi}{k}{\rm Im}\,{\rm Sp}(\rho f(0)).
\eeq
Here $k$ is the relative wave vector of colliding particles, and
$\rho$ is their spin density matrix.

Generally, the forward scattering amplitude may be represented in
terms of S-matrix elements. Let us consider an elastic scattering
of two particles with spins $s$ and $I$. The sum of the channel
spin $F$ (${\bf F}={\bf s}+{\bf I}$) and the relative orbital
momentum $l$ in an entrance channel gives the total angular
momentum $J$ (${\bf J}={\bf F}+{\bf l}$). In an exit channel the
channel spin $F'$ and the relative orbital momentum $l'$ can
differ from $F$ and $l$, respectively, provided the rules of
angular momenta summation ${\bf J}={\bf F'}+{\bf l'}$ and
${\bf F'}={\bf s}+{\bf I}$ are satisfied. Thus, such transition is
described by the S-matrix element $S_J(lF\to l'F')$. If TRI holds,
the S-matrix should be symmetric:
$S_J(lF\to l'F')=S_J(l'F'\to lF)$. It can be shown that the terms
in the total cross section (\ref{1}) related with the correlations
$({\bf n}_s[{\bf n}_k\times{\bf n}_I])$ and
$({\bf n}_s[{\bf n}_k\times{\bf n}_I])({\bf n}_k{\bf n}_I)$ are
proportional to differences
\beq{2}
S_J(lF\to l'F')-S_J(l'F'\to lF).
\eeq
Clearly, such terms are nonzero only if TRI breaks. When a light
particle is scattered by a heavy particle, in particular, in
neutron-nucleus interaction, it is more convenient to use the
total angular momentum of the light particle $j$
(${\bf j}={\bf s}+{\bf l}$, ${\bf J}={\bf j}+{\bf I}$) instead of
the channel spin $F$.

The first five-fold correlation test of TRI was made in \cite{8}
in the interaction of 2 MeV polarized neutrons with aligned nuclei
$^{165}$Ho. A bound of $\sim 10^{-2}$ on the ratio of T-odd forces
to T-even ones in the effective nucleon-nucleon interaction was
obtained. A similar test in the interaction of polarized protons
with aligned deuterons is now under preparation at the cooler
synchrotron COSY at Julich \cite{9,10}.

Both three- and five-fold correlations were proposed to be studied
in the interaction of resonance p-wave neutrons with heavy nuclei.
Such tests of TRI have the advantage that the effects may
be enhanced in a p-wave resonance by the factor of $\sim 10^3$
\cite{11,12}. The reason is the smallness of a resonance width
and, hence, an increase of the time of T-odd forces action. Note,
that the description of the effect for slow neutrons is quite
simple because of only three partial waves participate in the
scattering. Namely, $lj$=s1/2, p1/2 and p3/2, where $l$ is a
neutron orbital momentum, and $j$ is its total angular momentum.
Thus all scattering effects for resonance neutrons are determined
by nine \mbox{S-matrix} elements $S_J(lj\to l'j')$. Let us
consider a total cross section of polarized neutron interaction
with nuclei, which may be both polarized and aligned.

To describe nuclear orientation we choose an axis $z$ along the
unit vector ${\bf n}_I$. Let $m$ be a projection of the nuclear
spin $I$ on the $z$ axis, and $n_m$ is a population of the
$m$-substate ($\sum_mn_m=1$). Thus, we define nuclear polarization
and alignment as
\beq{3}
p_1(I)=\frac{\av{m}}{I},\qquad
p_2(I)=\frac{3\av{m^2}-I(I+1)}{I(2I-1)},
\eeq
where $\av{m^k}=\sum_mm^kn_m$. In the case of pure alignment
$n_m=n_{-m}$, thus $p_1(I)=0$. Both parameters $p_1(I)$ and
$p_2(I)$ equal unity when only the substate with the maximal
projection $m=I$ is populated ($n_m=\delta_{mI}$). By the same way
the neutron polarization is defined by $p_1(s)=\av{\sigma}/s$,
where $\sigma$ is a projection of the neutron spin $s=1/2$ on the
$z$ axis along the unit vector ${\bf n}_s$.

The total cross section of the interaction of slow neutrons with
nuclei is of the form
\beq{4}
\begin{array}{l}
\sigma_{tot}=\sigma_0+a_1p_1(s)p_1(I)({\bf n}_s{\bf n}_I)+
a_2p_1(s)p_1(I)(3({\bf n}_s{\bf n}_k)({\bf n}_I{\bf n}_k)-
({\bf n}_s{\bf n}_I))+{}
\\[\bigskipamount]
\phantom{\sigma_{tot}=\sigma_0}+
a_3p_2(I)(3({\bf n}_k{\bf n}_I)^2-1)+{}
\\[\bigskipamount]
\phantom{\sigma_{tot}}+
b_1p_1(s)({\bf n}_s{\bf n}_k)+
b_2p_1(I)({\bf n}_k{\bf n}_I)+
b_3p_1(s)p_2(I)(3({\bf n}_s{\bf n}_I)({\bf n}_k{\bf n}_I)-
({\bf n}_s{\bf n}_k))+{}
\\[\bigskipamount]
\phantom{\sigma_{tot}}+
c_1p_1(s)p_1(I)({\bf n}_s[{\bf n}_k\times {\bf n}_I])+
c_2p_1(s)p_2(I)({\bf n}_s[{\bf n}_k\times {\bf n}_I])
({\bf n}_k{\bf n}_I).
\end{array}
\eeq
Here $\sigma_0$ is the total cross section for unoriented neutrons
and nuclei. It can be presented in terms of S-matrix elements, as
well as the quantities $a_i$, $b_i$ and $c_i$ (see \cite{4,13}).
The terms related with $b_i$ are P-odd, while for T-odd terms we
have
\beq{5}
c_1=\frac{2\pi}{k^2}\sum_jC^{Jj}_1
{\rm Im}\left(S_J(0\frac{1}{2}\to 1j)-S_J(1j\to
0\frac{1}{2})\right),
\eeq
\beq{6}
c_2=\frac{2\pi}{k^2}C^J_2
{\rm Im}\left(S_J(1\frac{1}{2}\to 1\frac{3}{2})-
S_J(1\frac{3}{2}\to 1\frac{1}{2})\right),
\eeq
where $C^{Jj}_1$ and $C^J_2$ are numerical coefficients of the
unit scale. The three-fold  correlation arises from the asymmetry
of scattering from s1/2-wave to p1/2- and p3/2-waves and vice
versa. It is P-odd as the transitions between s- and p-waves are
parity violating. The five-fold correlation tests the equality of
the transition rates from p1/2-wave to p3/2-one and vice versa,
thus it is P-even. Clearly, both correlation should be studied in
p-wave resonances to maximize p-wave contribution to the
scattering. There exist additional possibilities to test TRI using
neutron spin rotation in transmission through polarized and
aligned targets (see, e.g., \cite{4,13}).

Experimental setup for development of research technique
for investigation  of TRI violation by studying  three-fold
and five-fold correlations  are being constructed now  at neutron
beam of pulsed reactor IBR-30 (JINR). The setup will include
well known neutron polarizer \cite{14} , neutron analyzer
(constructed in ITEP \cite{15}), system fo precise control and
adjustment of neutron spin and oriented nuclear targe. The
description of setup will be publised elsewhere.

\section{Dynamic nuclear alignment (DNA) method}

Let us consider a nucleus with a spin $I\geq 1$ and quadrupolar
moment $Q$ which is acted upon by an axial symmetric electric
field gradient (EFG) directed along an axis~$z$. An interaction of
$Q$ with EFG results in a set of  $(2I+1)/2$ sublevels (we assume
for simplicity that $I$ takes half-integer values). Each of them
is a degenerated doublet of substates with projections $\pm m$ of
spin $I$ on the axis $z$. The energy splitting of sublevels is
determined by a  parameter which is proportional both to nuclear
quadrupolar moment $Q$ and  EFG value. In the case under
consideration the energy differences between sublevels are equal
to $a$, $2a$, $3a$\ldots (from sublevel with $m=\pm 1/2$). One can
observes the signals of nuclear quadrupolar resonance (NQR) at
frequencies determined by the energy splitting. Their intensities
depend on substate populations $n_m$ and define the value of nuclear
alignment .

If the spins are in equilibrium at the temperature $T_0$, the
distribution $n_m$ over substates  is given by the Boltzman law.
For the case of $I=3/2$ the quadrupolar spectrum falls into two
sublevels with $m=\pm 1/2$ and $\pm 3/2$ separated by the energy
$a$. Then, $n_2/n_1=\exp(-a/kT)$, and for $a/h\sim 100$~MHz (for
a typical value of $a$ for heavy nuclei)
and $T_0=0.5$~K the equilibrium value of an alignment is
$p_2(I)=4.9\times 10^{-3}$. To obtain a higher nuclear alignment
the method of dynamic nuclear alignment (DNA) is proposed.

DNA method is similar to dynamic nuclear polarization (DNP)
method. However,  in the case of DNA there is no need in an
external magnetic field. The idea is following. Ground states of
paramagnetic ions with electron spin $S\geq 1$ are being
split in an electric field of crystal in the same way as the
states of quadrupolar nuclei. This splitting results from the
interaction of quadrupolar (and higher order) moment of an electron
shell of paramagnetic ion with EFG. The energy differences between
sublevels $h\Delta_0$ (where $\Delta_0$ is a frequency of electron
paramagnetic  resonance - EPR in zero magnetic field) may be 
of several orders of magnitude more
than those for nuclear quadrupolar splitting, and $\Delta_0$ may be
of order of tens GHz. Taking, for example, $S=3/2$,
$\Delta_0=50$~GHz  and T=0.3~K, we obtain completely aligned
electron spins $p_2(S)=1$ . One should emphasize that in this case
there is no necessity in an external magnetic field, and the
direction of quantization axis $z$ coincides with the main axis of
the crystal electric field. The dynamic nuclear alignment method
is based on the transmission of high alignment of electron spins
of paramagnetic admixture to the nuclei of the basic crystal
lattice.

As in the case of DNP  it may be realized by saturating
irradiation of a target by microwaves on a frequency
$\Delta_0+\delta$ near the resonance frequency of
paramagnetic ions. The  decreasing of spin temperature due to the
shift $\delta$ from the precise resonance frequency (which
corresponds to the center of EPR line in zero magnetic field)
leads to the decreasing of spin temperature $T_d$ of the electron
dipole-dipole subsystem
\beq{7}
\frac{T_0}{T_d}\simeq\frac{\Delta_0}{2\omega_L}.
\eeq
Here $\omega_L$ is a parameter of electron spin-spin interaction
which is of order of EPR linewidth at zero magnetic field (typical
values of $\omega_L$ are 100-300~MHz). As a consequence of
electron-nucleus dipole interaction this temperature is
transmitted to the spin subsystem of quadrupole nuclei. Thus, an
enhancement of nuclear alignment $p_2(I)$ arises by a factor of
$T_0/T_d\sim 10^2-10^3$ both in the cases of low  energy
($\delta<0$) and high energy ($\delta>0$) sublevels . The change 
of sign in $\delta$ can lead to change sign in $p_2$.
The values 0.4-0.8 for
nuclear alignment parameter $p_2(I)$ can be obtained.
Besides, a further nuclear alignment can be provided by "solid
effect" which can be used when EPR linewidth of paramagnetic
admixture is less than NQR linewidth . In this case it is possible
to obtain a full transmission of electron alignment to nuclei.

Theory of dynamic cooling and solid-effect is described in
\cite{16,17}. The decreasing of spin temperature at zero magnetic
field by the factor of 10$^2$ was observed in \cite{18} with the
use of Cr$^{3+}$ ions in rutile crystal (TiO$_2$) at 1.7~K and
microwave frequency $\Delta_0=43$~GHz. The method of dynamic
cooling was realized. This work did not involve nuclear alignment.
However, according our estimation an alignment of quadrupolar
nuclei included in a crystal at 0.3~K would be 0.5. The value of
nuclear alignment can be further increased by lowering of the
temperature and increasing of the pumping frequency.

A requirement on precise crystal orientation relative to the fixed
direction in the case of DNA is less strict than in the case of
DNP. Some misalignment would result in the lowering of an
alignment. However, it would not influence EPR linewidth which is
constant in the absence of magnetic field. So a misalignment of
the order of 20$^0$ is acceptable. This allows to produce a sample
with a large volume using a set of small crystals.

The main problem in realization of DNA method is a choosing of the
appropriate sample which has to fulfill the following criteria:

1)~high content of nuclei of interest,

2)~high quadrupolar energy splitting of nuclear  sublevels (NQR
frequency should be more than 30~MHz),

3)~high energy splitting of sublevel of paramagnetic admixture
(EPR frequency at zero magnetic field should be more than 30~GHz),

4)~the absence of the disoriented and nonequivalent sites of
quadrupoare nuclei and paramagnetic ions in crystal.

The last requirement is significant for achievement of an
acceptable degree of orientation along a separated axis.

\section{Summary}

From above consideration it is clear that any nuclei with spin
$I\geq 1$ having low lying \mbox{p-wave} resonances are good
candidates for TRI tests with aligned nuclei. Note, that about two
tens nuclei with low energy p-wave resonances were involved in P
violation experiments. However, only eight of them have quadrupolar
moment . From our point of view the more appropriate nuclei among
them are $^{35}$Cl, $^{81}$Br and $^{139}$La.

It should be pointed out that aligned target is of interest not
only for TRI violation experiment.
An alignment of deformed nuclei allows to study the deformation
effects in nuclear reactions. Up to now, such effects were
investigated only in the cross section of the elastic scattering
of neutrons by the aligned nuclei $^{59}$Co \cite{19} and
$^{165}$Ho \cite{20}. Angular correlations of secondary radiation
provide also the worth information on spin dependent reaction
amplitudes. For example, the energy and spin dependence of fission
amplitudes was recently obtained in the measurements of fission
fragment angular distributions in the fission of aligned nuclei
$^{235}$U by resonance neutrons \cite{21}.

In conclusion  we shall exemplify some materials containing
quadrupolar nuclei and doped with paramagnetic admixtures which
have large splitting of levels in EFG of crystal  (the NQR
frequencies for this materials were not measured): ScO$_2$  doped
with ion Cr$^{3+}$ (quadrupolar nucleus  $^{45}$Sc, I= 7/2,
$\Delta_0=70$~GHz ), Ga$_2$O$_3$ doped with ion Cr$^{3+}$
(quadrupolar nuclei $^{69}$Ga, $^{71}$Ga, I=3/2,
$\Delta_0=36$~GHz), Hf$_2$O$_3$ doped with ion Fe$^{3+}$
(quadrupolar nucleus $^{177}$Hf, I=7/2, $\Delta_0=60$~GHz) and
LiNbO$_3$ doped with ion Cr$^{3+}$ (quadrupolar nucleus $^{93}$Nb,
I=9/2, $\Delta_0=30$~GHz). This list is preliminary and can be
significantly extended.
\bigskip

This work was supported by the ISTC (grant N 608) and RFBR (grant
N 96-15-96548).

\end{document}